\input harvmac

\let\includefigures=\iftrue
\let\useblackboard=\iftrue
\newfam\black

%Figure Stuff
\includefigures
\message{If you do not have epsf.tex (to include figures),}
\message{change the option at the top of the tex file.}
\input epsf
\def\figin{\epsfcheck\figin}\def\figins{\epsfcheck\figins}
\def\epsfcheck{\ifx\epsfbox\UnDeFiNeD
\message{(NO epsf.tex, FIGURES WILL BE IGNORED)}
\gdef\figin##1{\vskip2in}\gdef\figins##1{\hskip.5in}% blank space instead
\else\message{(FIGURES WILL BE INCLUDED)}%
\gdef\figin##1{##1}\gdef\figins##1{##1}\fi}
\def\DefWarn#1{}
\def\figinsert{\goodbreak\midinsert}
\def\ifig#1#2#3{\DefWarn#1\xdef#1{fig.~\the\figno}
\writedef{#1\leftbracket fig.\noexpand~\the\figno}%
\figinsert\figin{\centerline{#3}}\medskip\centerline{\vbox{
\baselineskip12pt\advance\hsize by -1truein
\noindent\footnotefont{\bf Fig.~\the\figno:} #2}}
\bigskip\endinsert\global\advance\figno by1}
%%%
\else
\def\ifig#1#2#3{\xdef#1{fig.~\the\figno}
\writedef{#1\leftbracket fig.\noexpand~\the\figno}%
%\figinsert\figin{\centerline{#3}}\medskip
%\centerline{\vbox{\baselineskip12pt
%\advance\hsize by -1truein\noindent
%\footnotefont{\bf Fig.~\the\figno:} #2}}
%\bigskip\endinsert
\global\advance\figno by1}
\fi
%

%%%%new definitions
\def\xtp{{\tilde x^+}}
\def\xt{{\tilde x}}
\def\xtm{{\tilde x^-}}
\def\chit{{\tilde \chi}}

\def\yp{{y^+}}
\def\ym{{y^-}}
\def\chih{{\hat \chi}}
\def\pc{p^{\chi}}
\def\lslash{{L \over 2 \pi}}
%%%%

\def\CJ{{\cal J}}

\def\ap{\alpha'}
\def\si{{\sigma^1}}
\def\sii{{\sigma^2}}

\def\K3{{\bf K3}}
\def\journal#1&#2(#3){\unskip, \sl #1\ \bf #2 \rm(19#3) }
\def\andjournal#1&#2(#3){\sl #1~\bf #2 \rm (19#3) }

\def\hat{\widehat}

\def\tilde{\widetilde}

\def\frac#1#2{{#1\over#2}}

\def\half{\frac12}

\def\inbar{\,\vrule height1.5ex width.4pt depth0pt}
\def\IC{\relax\hbox{$\inbar\kern-.3em{\rm C}$}}
\def\IR{\relax{\rm I\kern-.18em R}}
\def\IP{\relax{\rm I\kern-.18em P}}

%
%%%%%%%%%%%%%%%%%%%%%%%%%%%%%%%%%%%%
%

%\def\ap#1#2#3{Ann. Phys. {\bf #1} (#2) #3}

%
\catcode`\@=11
\def\slash#1{\mathord{\mathpalette\c@ncel{#1}}}
\overfullrule=0pt

\def\underrel#1\over#2{\mathrel{\mathop{\kern\z@#1}\limits_{#2}}}

\catcode`\@=12

%%%%%%%%%%%%%%%%%%%%%%%%%%%%%%%%%%%%%%%%%%%%%%%%%%%%%%%%%%%%%%

%

\def \sinh{{\rm sinh}}
\def \cosh{{\rm cosh}}

\def\exp{{\rm exp}}

%%%%%%%%%%%%%%%%%%%%%%%%%%%%%%%%%%%%%%%%%%%%%%%%%%%%%%%%%%%%%%
% new defs:

\def\bX{{\bf X}}
\def\bG{{\bf G}}

\def \p {\partial}

\def\IL{\relax{\rm I\kern-.18em L}}
\def\IH{\relax{\rm I\kern-.18em H}}
\def\IR{\relax{\rm I\kern-.18em R}}
\def\IC{\relax\hbox{$\inbar\kern-.3em{\rm C}$}}
\def\IZ{{\bf Z}}

\def\CU{{\cal U}}

%\def\IZ{{\bf Z}}
%\def\IR{{\bf R}}

%%%%%%%%%%%%%%%%%%%%%%%%%%%%

%%%% you need these macros:
%%%%%
%%%%%

%%% MACROS FOR BOX BOUNDARY CONDS
%%% FROM KAWAI ET AL

\def\makeblankbox#1#2{\hbox{\lower\dp0\vbox{\hidehrule{#1}{#2}%
   \kern -#1% overlap rules
   \hbox to \wd0{\hidevrule{#1}{#2}%
      \raise\ht0\vbox to #1{}% vrule height
      \lower\dp0\vtop to #1{}% vrule depth
      \hfil\hidevrule{#2}{#1}}%
   \kern-#1\hidehrule{#2}{#1}}}%
}%
\def\hidehrule#1#2{\kern-#1\hrule height#1 depth#2 \kern-#2}%
\def\hidevrule#1#2{\kern-#1{\dimen0=#1\advance\dimen0 by #2\vrule
    width\dimen0}\kern-#2}%
\def\openbox{\ht0=1.2mm \dp0=1.2mm \wd0=2.4mm  \raise 2.75pt
\makeblankbox {.25pt} {.25pt}  }

\def\bun#1/#2{\leavevmode
   \kern.1em \raise .5ex \hbox{\the\scriptfont0 #1}%
   \kern-.1em $/$%
   \kern-.15em \lower .25ex \hbox{\the\scriptfont0 #2}%
}

\def\opensquare{\ht0=3.4mm \dp0=3.4mm \wd0=6.8mm  \raise 2.7pt
\makeblankbox {.25pt} {.25pt}  }

%%%%%%%%%%%%%%%%%%%%%%%

\def\sector#1#2{\ {\scriptstyle #1}\hskip 1mm
\mathop{\opensquare}\limits_{\lower 1mm\hbox{$\scriptstyle#2$}}\hskip 1mm}

\def\tsector#1#2{\ {\scriptstyle #1}\hskip 1mm
\mathop{\opensquare}\limits_{\lower 1mm\hbox{$\scriptstyle#2$}}^\sim\hskip 1mm}
%%%
%%%

%%%%%%%%%%%%%%%%%%%%%%%%
%OTHER
%%%%%%%%%%%%%%%

\def\xp{{x^+}}
\def\xm{{x^-}}
%%%%%%%%%%%%%%%%%%%%%%%%%%%%

\def\ap{\alpha'}

\def\IZ{{\bf Z}}

%%%%%%%%%%%%%%%%%%%%%%%%%%%%%%%%%%%
%OTHER DEFINITIONS
%%%%%%%%%%%%%%%%%%%%%%%%%%%%%%%%%%%
\def\ozero{{\IR^{1,3} / \Gamma_0} }
\def\ol{{\IR^{1,3} / \Gamma_L}}

%%%%%%%%%%%%%%%%%%%%%%%%%%%%%%%%%%%%%%%%%%%%%%%%%%%

%\lms
\lref\lms{
H.~Liu, G.~Moore and N.~Seiberg,
``Strings in a time-dependent orbifold,''
arXiv:hep-th/0204168.
%%CITATION = HEP-TH 0204168;%%
}

\lref\lmsnew{
H.~Liu, G.~Moore and N.~Seiberg,
to appear at the same time as this paper.
}

%\HorowitzAP
\lref\HorowitzAP{ G.~T.~Horowitz and A.~R.~Steif, ``Singular
String Solutions With Nonsingular Initial Data,'' Phys.\ Lett.\ B
{\bf 258}, 91 (1991).
%%CITATION = PHLTA,B258,91;%%
}

%\HorowitzBV
\lref\HorowitzBV{ G.~T.~Horowitz and A.~R.~Steif, ``Space-Time
Singularities In String Theory,'' Phys.\ Rev.\ Lett.\  {\bf 64},
260 (1990); ``Strings In Strong Gravitational Fields,'' Phys.\
Rev.\ D {\bf 42}, 1950 (1990).}
%%CITATION = PRLTA,64,260;%%
%

%\KhouryBZ
\lref\KhouryBZ{ J.~Khoury, B.~A.~Ovrut, N.~Seiberg,
P.~J.~Steinhardt and N.~Turok, ``From big crunch to big bang,''
arXiv:hep-th/0108187;
%%CITATION = HEP-TH 0108187;%%
%\SeibergHR
N.~Seiberg, ``From big crunch to big bang - is it possible?,''
arXiv:hep-th/0201039.
%%CITATION = HEP-TH 0201039;%%
}

%\BalasubramanianRY
\lref\BalasubramanianRY{ V.~Balasubramanian, S.~F.~Hassan,
E.~Keski-Vakkuri and A.~Naqvi, ``A space-time orbifold: A toy
model for a cosmological singularity,'' arXiv:hep-th/0202187.
%%CITATION = HEP-TH 0202187;%%
}
%\GutperleAI
\lref\GutperleAI{ M.~Gutperle and A.~Strominger, ``Spacelike
branes,'' arXiv:hep-th/0202210.
%%CITATION = HEP-TH 0202210;%%
}
%\CornalbaFI
\lref\CornalbaFI{ L.~Cornalba and M.~S.~Costa, ``A New
Cosmological Scenario in String Theory,'' arXiv:hep-th/0203031.
%%CITATION = HEP-TH 0203031;%%
}
%\NekrasovKF
\lref\NekrasovKF{ N.~A.~Nekrasov, ``Milne universe, tachyons, and
quantum group,'' hep-th/0203112.
%%CITATION = HEP-TH 0203112;%%
}

%\SenNU
\lref\SenNU{
A.~Sen,
``Rolling Tachyon,''
arXiv:hep-th/0203211.
%%CITATION = HEP-TH 0203211;%%
}

%\HiscockJQ
\lref\HiscockJQ{ W.~A.~Hiscock, ``Quantized fields and chronology
protection,'' arXiv:gr-qc/0009061.
%%CITATION = GR-QC 0009061;%%
}

%\HawkingNK
\lref\HawkingNK{ S.~W.~Hawking, ``The Chronology protection
conjecture,'' Phys.\ Rev.\ D {\bf 46}, 603 (1992).
%%CITATION = PHRVA,D46,603;%%
}

%\HawkingPK
\lref\HawkingPK{ S.~W.~Hawking, ``The Chronology Protection
Conjecture,'' {\it Prepared for 6th Marcel Grossmann Meeting on
General Relativity (MG6), Kyoto, Japan, 23-29 Jun 1991}. }

%\BrodskyDE
\lref\BrodskyDE{ S.~J.~Brodsky, H.~C.~Pauli and S.~S.~Pinsky,
``Quantum chromodynamics and other field theories on the light
cone,'' Phys.\ Rept.\  {\bf 301}, 299 (1998)
[arXiv:hep-ph/9705477].
%%CITATION = HEP-PH 9705477;%%
}

%\lref\nullbranes{
%\Figueroa
\lref\Figueroa{
J.~Figueroa-O'Farrill and J.~Sim\'{o}n,
``Generalized supersymmetric fluxbranes,''
JHEP {\bf 0112}, 011 (2001)
[arXiv:hep-th/0110170]}
%%CITATION = HEP-TH 0110170;%%
%
%\SimonMA
\lref\SimonMA{
J.~Sim\'{o}n, ``The geometry of null rotation
identifications,'' arXiv:hep-th/0203201.}
%%CITATION = HEP-TH 0203201;%%

\lref\longpaper{H. Liu, G. Moore, and N. Seiberg, to appear.}

%\CornalbaNV
\lref\CornalbaNV{
L.~Cornalba, M.~S.~Costa and C.~Kounnas,
``A resolution of the cosmological singularity with orientifolds,''
arXiv:hep-th/0204261.
%%CITATION = HEP-TH 0204261;%%
}

\lref\candelasetal{
P.~Candelas, T.~H\"ubsch and R.~Schimmrigk,
``Relation Between The Weil-Petersson And Zamolodchikov Metrics,''
Nucl.\ Phys.\ B {\bf 329}, 583 (1990).
%%CITATION = NUPHA,B329,583;%%
}

\lref\sandwich{
H.~Bondi, F.~A.~E.~Pirani and I.~Robinson,
``Gravitational waves in general relativity III. Exact plane waves,''
Proc.\ Roy.\ Soc.\ {\bf A251} (1959) 519;
}

\lref\rindler{
W.~Rindler, ``Essential Relativity'', Revised {\it Second} Edition, Springer-Verlag, New York,
1977.
}

%\WittenGJ
\lref\bubbles{
E.~Witten,
``Instability Of The Kaluza-Klein Vacuum,''
Nucl.\ Phys.\ B {\bf 195}, 481 (1982).
%%CITATION = NUPHA,B195,481;%%
}

%\lref\otherbubbles{
\lref\dggh{
F.~Dowker, J.~P.~Gauntlett, G.~W.~Gibbons and G.~T.~Horowitz,
``Nucleation of $P$-Branes and Fundamental Strings,''
Phys.\ Rev.\ D {\bf 53}, 7115 (1996)
[arXiv:hep-th/9512154]}
%%CITATION = HEP-TH 9512154;%%
%
%\DowkerGB
\lref\DowkerGB{
F.~Dowker, J.~P.~Gauntlett, G.~W.~Gibbons and G.~T.~Horowitz,
``The Decay of magnetic fields in Kaluza-Klein theory,''
Phys.\ Rev.\ D {\bf 52}, 6929 (1995)
[arXiv:hep-th/9507143]}
%%CITATION = HEP-TH 9507143;%%
%
%\ShinkaiAS
\lref\ShinkaiAS{
H.~A.~Shinkai and T.~Shiromizu,
``Fate of Kaluza-Klein bubble,''
Phys.\ Rev.\ D {\bf 62}, 024010 (2000)
[arXiv:hep-th/0003066]}
%%CITATION = HEP-TH 0003066;%%
%
%\FabingerJD
\lref\FabingerJD{
M.~Fabinger and P.~Ho\v{r}ava,
``Casimir effect between world-branes in heterotic M-theory,''
Nucl.\ Phys.\ B {\bf 580}, 243 (2000)
[arXiv:hep-th/0002073]}
%%CITATION = HEP-TH 0002073;%%
%
%\HorowitzGN
\lref\HorowitzGN{
G.~T.~Horowitz and L.~Susskind,
``Bosonic M theory,''
J.\ Math.\ Phys.\  {\bf 42}, 3152 (2001)
[arXiv:hep-th/0012037]}
%%CITATION = HEP-TH 0012037;%%
%\DeAlwisKP
\lref\DeAlwisKP{
S.~P.~De Alwis and A.~T.~Flournoy,
``Closed string tachyons and semi-classical instabilities,''
arXiv:hep-th/0201185}
%%CITATION = HEP-TH 0201185;%%
%
%\BirminghamST
\lref\BirminghamST{
D.~Birmingham and M.~Rinaldi,
``Bubbles in anti-de Sitter space,''
arXiv:hep-th/0205246}
%%CITATION = HEP-TH 0205246;%%
%
%\BalasubramanianAM
\lref\BalasubramanianAM{
V.~Balasubramanian and S.~F.~Ross,
``The dual of nothing,''
arXiv:hep-th/0205290}
%%CITATION = HEP-TH 0205290;%%
%
\lref\HorowitzMaeda{
G.T. Horowitz and K. Maeda, to appear.
}

%\lref\otherbackgrounds{
%
%\AmatiSA
\lref\AmatiSA{
D.~Amati and C.~Klim\v{c}\'{\i}k,
``Nonperturbative Computation Of The Weyl Anomaly For A Class Of
Nontrivial Backgrounds,''
Phys.\ Lett.\ B {\bf 219}, 443 (1989).}
%%CITATION = PHLTA,B219,443;%%
%
%\deVegaNR
\lref\deVegaNR{
H.~J.~de Vega and N.~Sanchez,
``Space-Time Singularities In String Theory And String Propagation
Through Gravitational Shock Waves,''
Phys.\ Rev.\ Lett.\  {\bf 65}, 1517 (1990).}
%%CITATION = PRLTA,65,1517;%%
%
%\NappiKV
\lref\NappiKV{
C.~R.~Nappi and E.~Witten,
``A Closed, expanding universe in string theory,''
Phys.\ Lett.\ B {\bf 293}, 309 (1992)
[arXiv:hep-th/9206078].}
%%CITATION = HEP-TH 9206078;%%
%
%\NappiIE
\lref\NappiIE{
C.~R.~Nappi and E.~Witten,
``A WZW model based on a nonsemisimple group,''
Phys.\ Rev.\ Lett.\  {\bf 71}, 3751 (1993)
[arXiv:hep-th/9310112].}
%%CITATION = HEP-TH 9310112;%%
%
\lref\KiritsisJK{
E.~Kiritsis and C.~Kounnas,
``String Propagation In Gravitational Wave Backgrounds,''
Phys.\ Lett.\ B {\bf 320}, 264 (1994)
[Addendum-ibid.\ B {\bf 325}, 536 (1994)]
[arXiv:hep-th/9310202].}
%%CITATION = HEP-TH 9310202;%%
%
%\KiritsisIJ
\lref\KiritsisIJ{
E.~Kiritsis, C.~Kounnas and D.~L\"ust,
``Superstring gravitational wave backgrounds with space-time supersymmetry,''
Phys.\ Lett.\ B {\bf 331}, 321 (1994)
[arXiv:hep-th/9404114].}
%%CITATION = HEP-TH 9404114;%%

%
%\lref\newexamples{
\lref\SilversteinXN{
%\SilversteinXN
E.~Silverstein,
``(A)dS backgrounds from asymmetric orientifolds,''
\ [arXiv:hep-th/0106209].}
%%CITATION = HEP-TH 0106209;%%
%
%\KoganNN
\lref\KoganNN{
I.~I.~Kogan and N.~B.~Reis,
``H-branes and chiral strings,''
Int.\ J.\ Mod.\ Phys.\ A {\bf 16}, 4567 (2001)
[arXiv:hep-th/0107163].}
%%CITATION = HEP-TH 0107163;%%
%
%\AdamsSV
\lref\AdamsSV{
A.~Adams, J.~Polchinski and E.~Silverstein,
``Don't panic! Closed string tachyons in ALE space-times,''
JHEP {\bf 0110}, 029 (2001)
[arXiv:hep-th/0108075].}
%%CITATION = HEP-TH 0108075;%%}
%
%
%
%\BerglundAJ
\lref\BerglundAJ{
%\BerglundAJ
P.~Berglund, T.~H\"ubsch and D.~Minic,
``de Sitter spacetimes from warped compactifications of IIB string  theory,''
arXiv:hep-th/0112079.}
%%CITATION = HEP-TH 0112079;%%
%
%\GutperleAI
\lref\GutperleAI{
M.~Gutperle and A.~Strominger,
``Spacelike branes,''
JHEP {\bf 0204}, 018 (2002)
[arXiv:hep-th/0202210].}
%%CITATION = HEP-TH 0202210;%%
%
%\BuchelWF
\lref\BuchelWF{
A.~Buchel,
``Gauge / gravity correspondence in accelerating universe,''
arXiv:hep-th/0203041.}
%%CITATION = HEP-TH 0203041;%%
%
%\SenNU
\lref\SenNU{
A.~Sen,
``Rolling tachyon,''
JHEP {\bf 0204}, 048 (2002)
[arXiv:hep-th/0203211].}
%%CITATION = HEP-TH 0203211;%%
%
%\ChenYQ
\lref\ChenYQ{
C.~M.~Chen, D.~V.~Gal'tsov and M.~Gutperle,
``S-brane solutions in supergravity theories,''
arXiv:hep-th/0204071.}
%%CITATION = HEP-TH 0204071;%%
%

\lref\aharony{
O.~Aharony, M.~Fabinger, G.~Horowitz, and E.~Silverstein,
``Clean Time-Dependent String Backgrounds from Bubble Baths,''
arXiv:hep-th/0204158.}

%\KruczenskiAP
\lref\KruczenskiAP{
M.~Kruczenski, R.~C.~Myers and A.~W.~Peet,
``Supergravity S-branes,''
JHEP {\bf 0205}, 039 (2002)
[arXiv:hep-th/0204144].}
%%CITATION = HEP-TH 0204144;%%
%
%\ElitzurRT
\lref\ElitzurRT{
S.~Elitzur, A.~Giveon, D.~Kutasov and E.~Rabinovici,
``From big bang to big crunch and beyond,''
arXiv:hep-th/0204189.}
%%CITATION = HEP-TH 0204189;%%
%
%\CornalbaNV
\lref\CornalbaNV{
L.~Cornalba, M.~S.~Costa and C.~Kounnas,
``A resolution of the cosmological singularity with orientifolds,''
arXiv:hep-th/0204261.}
%%CITATION = HEP-TH 0204261;%%
%
%\CrapsII
\lref\CrapsII{
B.~Craps, D.~Kutasov and G.~Rajesh,
``String propagation in the presence of cosmological singularities,''
arXiv:hep-th/0205101.}
%%CITATION = HEP-TH 0205101;%%
%
%\RoyIK
\lref\RoyIK{
S.~Roy,
``On supergravity solutions of space-like Dp-branes,''
arXiv:hep-th/0205198.}
%%CITATION = HEP-TH 0205198;%%
%
%\KachruKX
\lref\KachruKX{
S.~Kachru and L.~McAllister,
``Bouncing brane cosmologies from warped string compactifications,''
arXiv:hep-th/0205209.}
%%CITATION = HEP-TH 0205209;%%
%
\lref\maloney{
A.~Maloney, E.~Silverstein, and A.~Strominger,
``de Sitter Space in Non-Critical String Theory,''
arXiv:hep-th/0205316.}

%\DegerIE
\lref\DegerIE{ N.~S.~Deger and A.~Kaya, ``Intersecting S-Brane
Solutions of D=11 Supergravity,'' arXiv:hep-th/0206057.
%%CITATION = HEP-TH 0206057;%%
}

%\LawrenceAJ
\lref\LawrenceAJ{
A.~Lawrence,
``On the instability of 3D null singularities,''
arXiv:hep-th/0205288.
%%CITATION = HEP-TH 0205288;%%
}

\lref\Tseytlin{
%\TseytlinSB
%\lref\TseytlinSB{
A.~A.~Tseytlin,
``Exact string solutions and duality,''
arXiv:hep-th/9407099;
%%CITATION = HEP-TH 9407099;%%
C.~Klim\v{c}\'{\i}k and A.~A.~Tseytlin, unpublished (1994);
A.~A.~Tseytlin, unpublished (2001).
}

%\lref\twistedcircle{

%\TseytlinEI
\lref\TseytlinEI{
A.~A.~Tseytlin,
``Melvin solution in string theory,''
Phys.\ Lett.\ B {\bf 346}, 55 (1995)
[arXiv:hep-th/9411198].
%%CITATION = HEP-TH 9411198;%%
}
%\RussoIK
\lref\RussoIK{
J.~G.~Russo and A.~A.~Tseytlin,
``Magnetic flux tube models in superstring theory,''
Nucl.\ Phys.\ B {\bf 461}, 131 (1996)
[arXiv:hep-th/9508068].}
%%CITATION = HEP-TH 9508068;%%
%
%\CostaNW
\lref\CostaNW{
M.~S.~Costa and M.~Gutperle,
``The Kaluza-Klein Melvin solution in M-theory,''
JHEP {\bf 0103}, 027 (2001)
[arXiv:hep-th/0012072].}
%%CITATION = HEP-TH 0012072;%%
%
%\GutperleMB
\lref\GutperleMB{
M.~Gutperle and A.~Strominger,
``Fluxbranes in string theory,''
JHEP {\bf 0106}, 035 (2001)
[arXiv:hep-th/0104136].}
%%CITATION = HEP-TH 0104136;%%
%
%\RussoTF
\lref\RussoTF{
J.~G.~Russo and A.~A.~Tseytlin,
``Magnetic backgrounds and tachyonic instabilities in closed superstring  theory and M-theory,''
Nucl.\ Phys.\ B {\bf 611}, 93 (2001)
[arXiv:hep-th/0104238].}
%%CITATION = HEP-TH 0104238;%%
%
%\MotlDJ
\lref\MotlDJ{
L.~Motl,
``Melvin matrix models,''
arXiv:hep-th/0107002.}
%%CITATION = HEP-TH 0107002;%%
%
%\TseytlinQB
\lref\TseytlinQB{
A.~A.~Tseytlin,
``Magnetic backgrounds and tachyonic instabilities in closed string  theory,''
arXiv:hep-th/0108140.}
%%CITATION = HEP-TH 0108140;%%
%
%\DavidVM
\lref\DavidVM{
J.~R.~David, M.~Gutperle, M.~Headrick and S.~Minwalla,
``Closed string tachyon condensation on twisted circles,''
JHEP {\bf 0202}, 041 (2002)
[arXiv:hep-th/0111212].}
%%CITATION = HEP-TH 0111212;%%

\lref\ppwaves{J.~Ehlers and W.~Kundt, ``Exact Solutions of the
Gravitational Field Equations,'' in ``Gravitation: an introduction to current research,''
ed. L.~Witten, New York, Wiley (1962). See also
C.~Misner, K.~S.~Thorne and J.~A.~Wheeler, ``Gravitation,''
San Francisco, W. H. Freeman (1973), Exercise 35.8.
}

\lref\rohm{R. Rohm, ``Spontaneous supersymmetry breaking in
supersymmetric string theories,''  Nucl.\ Phys.\ B {\bf 237}, 553 (1984). }

%\WittenHC
\lref\WittenHC{
E.~Witten,
``(2+1)-Dimensional Gravity As An Exactly Soluble System,''
Nucl.\ Phys.\ B {\bf 311}, 46 (1988).
%%CITATION = NUPHA,B311,46;%%
}

%\HorowitzMW
\lref\HorowitzMW{ G.~T.~Horowitz and J.~Polchinski, ``Instability
of spacelike and null orbifold singularities,''
arXiv:hep-th/0206228.
%%CITATION = HEP-TH 0206228;%%
}

%%%%%%%%%%%%%%%%%%%%%%%%%%%%%%%%%%%%%%%%%%%%%%%%
%\special{"%
%  gsave %
%  %Resolution 72 div dup neg scale %
%  %currentpoint translate % 50 rotate
%  gsave 80 -580 translate 50 rotate %
%  /Times-Roman findfont 180 scalefont setfont %
%  10 -1 0 { dup .03 mul .7 add setgray dup neg moveto (DRAFT) show } for %
%  .9 setgray 0 0 moveto (DRAFT) show %
% % 150 150 translate %
% % /Times-Roman findfont 80 scalefont setfont %
% % .95 setgray 2 setlinewidth 0 0 moveto (\draftbeer) false charpath stroke %
%  grestore} %
%%%%%%%%%%%%%%%%%%%%%%%%%%%%%%%%%%%%%%%%%%%%%%%%%%%%%%%%%%%%%%%

\Title{\vbox{\baselineskip12pt \hbox{hep-th/0206196}
\hbox{SLAC-PUB-9256}
\hbox{SU-ITP-02-24}
}}%
{\vbox{\centerline{On Smooth Time-Dependent Orbifolds}
\medskip
{\vbox{\centerline{ and Null Singularities} }}
}}

\smallskip
\centerline{Michal Fabinger and John McGreevy}
\bigskip

\centerline{ \sl Department of Physics and SLAC}
\centerline{ \sl Stanford University}
\centerline{ \sl Stanford, CA 94305/94309}

\bigskip
\bigskip
\bigskip
\bigskip

%\vglue .3cm

\leftskip 8mm  \rightskip 8mm \baselineskip14pt \noindent
%\tenrm
%{\tenbf \hskip 7mm Abstract.}

 We study string theory on a non-singular
time-dependent orbifold of flat space,
%%b
known as the `null-brane'.
%%e
The orbifold group, which involves only space-like
identifications, is obtained by a combined action of a null
Lorentz transformation and a constant shift in an extra direction.
In the limit where the shift goes to zero, the geometry of this
orbifold reproduces an orbifold with a light-like singularity,
which was recently studied by Liu, Moore and Seiberg
(hep-th/0204168). We find that the backreaction on the geometry
due to a test particle can be made arbitrarily small, and that
there are scattering processes which can be studied in the
approximation of a constant background. We quantize strings on
this orbifold and calculate the torus partition function. We
construct a basis of states on the smooth orbifold whose tree
level string interactions are nonsingular. We discuss the
existence of physical modes in the singular orbifold which resolve
the singularity. We also describe another way of making the
singular orbifold smooth which involves a sandwich pp-wave.

\leftskip 0mm  \rightskip 0mm
\Date{\hskip 8mm June 2002}

%\draftmode

%%%%%%%%%%%%%%%%%%%%%%%%%%%%%%%%%

\newsec{Introduction}

Most of our present knowledge of string theory
pertains to
time-independent backgrounds.
However, some of the most interesting questions
we would like to ask a theory of quantum gravity,
namely those related to cosmological singularities
and horizons,
belong fully to the realm of time-dependent spacetimes.
If we want to understand how string theory answers these questions,
we need to know
how to formulate string theory in such spacetimes.

This is in general a
difficult problem.
As a simple class of examples,
it seems natural to study time-dependent orbifolds of
flat Minkowski space by a
discrete subgroup of the Poincar\'e group
\refs{
\WittenHC \HorowitzAP \Tseytlin \KhouryBZ  \CornalbaFI
\NekrasovKF \Figueroa \SimonMA \lms - \LawrenceAJ}
 (see also \BalasubramanianRY).
Many such orbifolds contain closed timelike curves, which raise
unpleasant issues. Better in this regard is the model studied by
Liu, Moore and Seiberg \lms\ which is an orbifold by $\IZ$
generated by a parabolic element of $SO(1,2)$ and belongs to the
class of models described by Horowitz and Steif \HorowitzAP. The
orbifold has a light-like singularity and contains closed
light-like curves. It has a null Killing vector, which allows one
to use light-cone quantization. (See  \LawrenceAJ\ for a
discussion of the stability of the singularity in this orbifold.
Various other time-dependent backgrounds of string theory were
studied,  for example, in \refs{ \AmatiSA \HorowitzBV \deVegaNR
\NappiKV \NappiIE \KiritsisJK - \KiritsisIJ}, and more recently in
\refs{ \SilversteinXN \KoganNN \AdamsSV \BerglundAJ \GutperleAI
\BuchelWF \SenNU \ChenYQ  \KruczenskiAP \aharony \ElitzurRT
\CornalbaNV \CrapsII \RoyIK \KachruKX  \maloney - \DegerIE}.)

In the first part of this paper we consider string theory
on a very closely-related orbifold, which was recently discussed in
\refs{\Figueroa,\SimonMA}.
The generator of the orbifold group is a parabolic element of $SO(1,2)$
combined with a constant shift in a fourth direction.
Its main virtue is that the orbifold group has no fixed points,
and therefore the quotient space contains no singularities at all.
In the limit where the shift goes to zero we recover precisely
the orbifold of \lms.

Studying this singular limit
provides a new perspective on the
interesting null singularity of \lms.
Perhaps more significantly, the orbifold with the shift
provides a time-dependent string background which has
a free world-sheet description, and in which
the backreaction is under control.
The ability to study the issue of time-dependence separately
from complications raised by the presence of singularities
is likely to be quite useful.

For example, despite the solvable and smooth nature of
the world-sheet theory for this model,
it is still not clear how to do string calculations
to an arbitrary loop order in the covariant gauge.
The orbifold does not have any useful Euclidean continuation,
so one is forced to do computations in Lorentzian signature of space-time
and therefore also in Lorentzian signature on the world-sheet.
Riemann surfaces of genus different from one, however,
do not admit any smooth Lorentzian metric, posing an obvious
difficulty for covariant calculations.

In the second part of the paper, we ask whether in the
$\IR^{1,3} \times \IR^6$ model of Liu, Moore and Seiberg \lms,
specifying the initial data for the geometry
only on a slice of a constant light-cone time determines that
 there will be a singularity in the future (or in the past).
We find that the answer to this question is negative
and we present everywhere non-singular
gravitational solutions whose light-cone-time past
exactly coincides with that of the singular orbifold
of Liu, Moore and Seiberg \lms. In addition to the light-cone-time
past, one can, if desired, make also late light-cone-time future
coincide with that
\lms.

While this work was in progress we learned
that a similar study was being made \lmsnew,
and that the smooth time-dependent orbifold
of Minkowski space was also suggested
by Joe Polchinski and Eva Silverstein.

The organization of this paper is as follows.  In \S2 we describe the
classical geometry of the nonsingular orbifold and discuss some
sets of coordinates which will be useful.  In \S3 we probe the
geometry with test masses and show that it does not
collapse gravitationally.  In \S4 we consider the wavefunctions of a
scalar field
in this background.  In \S5 we quantize strings on the orbifold in
light-cone gauge, and calculate the partition function.
In \S6 we study tree level amplitudes on the orbifold and
show that there is a basis of wavefunctions in which they
are nonsingular.
In \S7 we
study the mode of the singular orbifold which turns on the shift.
In \S8 we resolve the singularity of \lms\ with a sandwich wave.
In \S9 we conclude.

%%%%%%%%%%%%%%%%%%%%%%%%%%%%%%%%%%%%
\newsec{Classical Geometry}

The geometry we will study is a $\IZ$ orbifold of flat Minkowski space $\IR^{1,3}$
(times $\IR^{6}$ or $ \IR^{22}$, depending on whether we want to consider superstrings or bosonic strings).
In terms of coordinates
\eqn\coord{\xp = {x^0 + x^1\over \sqrt{2}},\quad  \xm = {x^0 - x^1\over \sqrt{2}}, \quad x = x^2,\quad  \chi = x^3,}
the metric is
\eqn\metric{ds^2 = -2 \xp\xm + dx^2 + d\chi^2}
We will write the generator of the orbifold group $\Gamma_L$ as
 \eqn\generator{{\sl g}_L = \exp\left( i v J \right) \ \exp\left(i L p^\chi \right), \quad J \equiv {1 \over
\sqrt{2}}J^{x^0x} + {1\over \sqrt{2}}J^{x^1x}}
This corresponds to a composition of a null Lorentz transformation of the $(\xp , x, \xm)$ subspace and a translation by
$L$ in the $\chi$-direction. In terms of the spacetime coordinates, ${\sl g}_L$  acts
as
\eqn\orbifoldaction{ \pmatrix{ x^+ \cr x^{~}\cr x^-\cr \chi \cr} \quad \rightarrow\quad
\pmatrix{ x^+ \cr x + v x^+  \cr
x^- + v x + \half v^2 x^+ \cr \chi  + L\cr}
}
For $L=0$ the orbifold becomes the orbifold studied by Liu, Moore and Seiberg \lms, which is
singular at $\xp=0$. For non-zero $L$
this orbifold is completely smooth and does not have any closed time-like or space-like curves.
The spacetime interval ${( n v \xp)^2 + L^2}$ between points identified by the $n$-th
 power of the orbifold group generator \generator\ is strictly positive.
 The $\IR^{1,3} / \Gamma_{L=0}$ orbifold has the property that every spacetime
 point with $\xp>0$ is in the causal future of every point with $\xp <0$ \lms.
 This is no longer true for $L \ne 0$.

The $\ol$ orbifold in general preserves the subgroup of the four-dimensional Poincar\'e symmetry group generated by
$p^\chi, J$, and $p^+=-p_-$.
For non-zero $L$, the
topology of the spacetime is simply $\IR^3 \times S^1$, which can be made manifest
by defining a new set of coordinates $\xtp, \xt, \xtm, \chit$ by
 \eqn\tcoord{ \eqalign{
 \xtp  &=   \xp \cr
 \xt \ \  & =    x - { \chi \over L} v \xp \cr
\xtm & =    \xm - {\chi \over L} v x + \half {\chi^2 \over L^2} v^2 \xp \cr
 \chit \ \  & = {\chi \over L}, \cr}}
in terms of which the orbifold action  becomes
\eqn\tid{(\xtp,\xt,\xtm,\chit)\quad\rightarrow\quad(\xtp,\xt,\xtm,\chit+1).}
The map between $(\xp,x,\xm,\chi)$ and $(\xtp,\xt,\xtm,\chit)$ is everywhere smooth and one-to-one, and can be easily inverted:
 \eqn\tcoordinv{ \eqalign{
 x^+  &=   \xtp \cr
 x \ \ & =    \xt + { \chit} v \xtp \cr
 x^- & =    \xtm + { \chit} v \xt + \half { \chit^2} v^2 \xtp \cr
 \chi \ \ & = {L \chit}, \cr}}
The metric  now becomes
%%
%\eqn\tmetric{ds^2 = -2 d\xtp d\xtm + d\xt^2 + \left[ 1+ \left( {v\xtp \over L}\right)^2 \right] d\chi^2
%- {2v \over L} (\xt d\xtp + \xtp d\xt) d\chi}
%%
%One can also define  $\chit = \chi / L$, to obtain identifications
%%
%\eqn\ttid{(\xtp,\xt,\xtm,\chit)\quad\rightarrow\quad(\xtp,\xt,\xtm,\chit+1).}
%
%and metric
%
\eqn\tmetric{ds^2 = -2 d\xtp d\xtm + d\xt^2 + \left[ L^2 + ( v\xtp )^2 \right] d\chit^2
+ 2v \ (\xtp d\xt - \xt d\xtp)   d\chit.}
which is non-degenerate everywhere.

%%b Two pi replaced by v.
Away from $\xp = 0$ one can define another useful set of coordinates,
 \eqn\ycoordinates{ \eqalign{ y^+  & = x^+   \cr
  y \ \   & =  {x\over x^+}   \cr
  y^- &  = x^- -  \half{x^2\over x^+} \cr
  \psi \ \ & = \chi - {L\over v} {x \over  x^+}. \cr}}
The corresponding identifications and metric are
 \eqn\yidentifications{(y^+, y, y^-, \psi) \sim (y^+, y+v , y^-, \psi) ,}
 \eqn\ymetric{\eqalign{
 ds^2 & = -2dy^+ dy^- + \left((y^+)^2 + {L^2 \over v^2}\right) dy^2 + d \psi^2 + {L\over
\pi}dyd\psi .
 \cr}}
 The definition of the $y$-coordinates in \ycoordinates\ is the same as in \lms.
%%e

%%%%%%%%%%%%%%%%%%%%%%%%%%%%original (longer) version %%%%%%%%%%%%%%%%%%%%%%%%%%%%%%%%%%%
%The (Hausdorff) topology of the spacetime is simply $\IR^3 \times S^1$, which can be made manifest
%by defining a new set coordinates $(\xtp, \xt, \xtm, \chi)$,
%%
%%\eqn\tcoord2{
%%\pmatrix{ x^+ \cr x \cr x^-\cr \chi \cr} \quad = \quad \pmatrix{ \xtp \cr \xt + {\chi \over L}
%%v \xtp  \cr \xtm + {\chi \over L} v \xt + \half {\chi^2 \over L^2} v^2 \xtp \cr \chi}}
%%
%%
% \eqn\tcoord{ \eqalign{
% x^+  &=   \xtp   \cr
% x  & =    \xt + {\chi \over L} v \xtp \cr
% x^- & =    \xtm + {\chi \over L} v \xt + \half {\chi^2 \over L^2} v^2 \xtp \cr
% \chi & = \chi, \cr}}
%%
%in terms of which the identifications become
%%
%\eqn\tid{(\xtp,\xt,\xtm,\chi)\quad\rightarrow\quad(\xtp,\xt,\xtm,\chi+L).}
%%
%The map between $(\xp,x,\xm,\chi)$ and $(\xtp,\xt,\xtm,\chi)$ is everywhere smooth and one-to-one.
%The metric can be now expressed as
%%
%\eqn\tmetric{ds^2 = -2 d\xtp d\xtm + d\xt^2 + \left[ 1+ \left( {v\xtp \over L}\right)^2 \right] d\chi^2
%- {2v \over L} (\xt d\xtp + \xtp d\xt) d\chi}
%%
%One can also define  $\chit = \chi / L$, to obtain identifications
%%
%\eqn\ttid{(\xtp,\xt,\xtm,\chit)\quad\rightarrow\quad(\xtp,\xt,\xtm,\chit+1).}
%%
%and metric
%%
%\eqn\ttmetric{ds^2 = -2 d\xtp d\xtm + d\xt^2 + \left[ L^2 + ( v\xtp )^2 \right] d\chi^2
%- 2v \ (\xt d\xtp + \xtp d\xt)   d\chi.}
%%
%Note that the metric \tmetric, \ttmetric\  is non-degenerate everywhere.

\newsec{Backreaction of particles on the geometry}

One of the first questions that arises when one considers time-dependent orbifolds is whether
the presence of a single particle causes the spacetime to gravitationally collapse. Placing one such particle
of rest mass $m$ (say $m \ne 0$) in the orbifold corresponds to adding to the universal
covering space  an infinite number of
particles (the original one plus its images) which are boosted with respect to each other. Since the boost
of distant particles goes to infinity, one might worry that the mass of a finite number of them might
 be larger than the corresponding Schwarzschild radius (which would be a clear sign of a large backreaction).

We will see that if $L \ne 0$ this does not happen here, provided $m$ is not too large. Suppose we work in an
inertial frame in which the `original particle' is at rest. At any time $x^0$ the distance to its $n$-th
image will be no smaller than $nL$, i.e. it grows at least linearly with $n$. The velocity of the $n$-th image is
%
%\eqn\velocity{{\sl v}_n= {\sqrt{8n^2v^2 + n^4v^4}\over{4+n^2v^2}},}
\eqn\velocity{{\sl v}_n= {nv\over{4+n^2v^2}}\sqrt{8 + n^2v^2},}
and corresponds to energy
\eqn\energy{E_n = m \gamma_n= {m\over \sqrt{1- {\sl v}_n^2} } = m
\left(1 + {1\over 4} n^2 v^2 \right) \sim {1\over 4}m\ \! n^2 v^2
.}
%
%
%%b This was confusing, because the energy computed is not the cms energy.
%
%
%Therefore the total energy of the first $2n$ images grows like $n^3$. Taking into account more and more images, the corresponding
%Schwarzschild radius will grow like $(n^3)^{1/D-3} = n^{3/D-3}$ in spacetime dimension $D$.
%In order for our lower bound $nL$ on the size of the region containing the first $2n$
%images to grow faster than the Schwarzschild radius we need $D>6$. This is of course satisfied in our case where the universal covering
%space is $\IR^{1,3}$ times $\IR^{6}$. As a result, the gravitational backreaction of a particle of small
%enough mass  on the geometry of the orbifold (with large enough $L$) will be small, and in particular,
%it will not cause a gravitational collapse.
%(Of course, we cannot repeat the same argument for the orbifold with
%$L = 0$.)
%
Therefore the total energy of the first $2n$ images  grows like
$n^3$. This energy is not the center-of-mass frame energy of the
first $2n$ images (which is actually smaller), but even if it were,
the corresponding Schwarzschild radius would not grow faster than
$(n^3)^{1/(10-3)} = n^{3/7}$, since we work in ten
dimensions.\foot{The
 center-of-mass frame energy grows like $n^2$, leading
 to a gravitational radius of order $n^{2/7}$,
 as discussed in detail in
\HorowitzMW.}

For non-zero $L$, this is still grows slower than the smallest
size $nL$ of the region containing the first $2n$ images. As a
result,  adding a  particle of small enough energy to the
null-brane does not cause a gravitational collapse. (Clearly, we
cannot make a similar statement for the parabolic orbifold, where
$L=0$).
%%e

For massless particles, we need to modify the argument just very slightly. Pick an inertial frame in which the particle
 has an energy $E_0$.
 The images will be boosted  in some direction by the same gamma-factor $\gamma_n$ as in \energy .
This means that $E_n \le (1+v) \gamma_n\ \! E_0$, leading to the same conclusion as before.

A very similar kind of reasoning shows that for small enough energies
(and large enough $L$),
scattering amplitudes of approximately localized particles (or strings)
are not affected much by the image particles,
and therefore they are
inherited from the covering space with just small corrections.
This means that there exist scattering amplitudes
for which it does make sense to work in the approximation of a fixed background, as we do in \S6.
Of course, for large enough energies
this approximation will break down, just like in any other spacetime.

%%%%%%%%%%%%%%%%%%%%%%%%%%%%%%%%%%%%%%%
\newsec{Scalar particle wavefunctions on the smooth $\ol$ orbifold}
%%%%%%%%%%%%%%%%%%%%%%%%%%%%%%%%%%%%%%%

First quantized wavefunctions of scalar particles in the $\IR^{1,3} / \Gamma_{L \ne 0}$ orbifold
are a close analog of the wavefunctions for the $L=0$ case \lms. The wave equation
for scalars is
 \eqn\scalareq{ \bigl[-2 {\p \over \p x^+}{\p \over \p x^-}+ ({\p
 \over \p x })^2   +({\p \over \p \chi })^2\bigr] \Psi = m^2 \Psi .}
The orbifold group generator \generator\ can be expressed
as\foot{From now on we will set $v=2\pi$ for simplicity.}
 \eqn\gen{\eqalign{{
 \CU( {\sl g}_L)} = \exp(2\pi i \hat J)\ \exp( i L \hat p^{\chi}), \qquad \qquad \qquad
  \cr
\hat J = \hat x^+ \hat p - \hat x \hat p^+ = -i \bigl( x^+{\p
\over \p x} + x {\p \over \p x^-}\bigr)
,\quad \hat p^+ = -i {\p \over \p \xm }
,\quad \hat p^{\chi} = -i {\p \over \p \chi }.\cr }}
Because $\CU( {\sl g}_L)$ commutes with $\hat p^+, \hat J$, and $\hat p^\chi$
we may use the corresponding eigenvalues to label the wavefunctions.
Up to the case of $p^+ = 0 $, they have the form
 \eqn\wavefunctions{\eqalign{ \Psi_{p^+,J, p^\chi}  = &\sqrt{p^+\over ix^+}\
 \exp \left[-ip^+x^--i{{m^2 + p^2_\chi}\over 2p^+} x^+ + i{ p^+
 \over 2 x^+}(x-\xi)^2 + i p^\chi \chi\right] \cr
& = \int_{-\infty}^\infty {dp  \over \sqrt{2 \pi }}\
 e^{-ip\xi}  \phi_{p^+,p,p^\chi}(x^+,x^-, x, \chi) \cr
 }}
where $ \xi = - J/p^+$ and
\eqn\phidef{\phi_{p^+,p, p^\chi}(x^+,x,
 x^-, \chi)=\exp\left(-ip^+x^- -ip^-x^+ +i px +i p^\chi \chi \right), \ p^-=
 {m^2 + p^2 + p^2_\chi \over 2p^+}.}
In the limit $x^+ \to 0$ the wavefunctions \wavefunctions\ become
 \eqn\psili{\lim_{x^+ \to 0}\Psi_{p^+,J,p^\chi} (x^+ ,x, x^-, \chi) =
 \sqrt{ 2\pi} e^{-ip^+x^- + i p^\chi \chi}\delta(x-\xi), \quad \xi=-J/p^+.}
In the $L=0$ case  a similar statement
implied \lms\ that the wavefunctions of definite
$p^+$ were supported at $x^+=0$ on the lattice $x\in {1\over p^+}
\IZ$. Here the situation is just slightly more complicated.
In terms of \wavefunctions\ taking the orbifold corresponds to
making
$2\pi J + L p^\chi$ quantized:
\eqn\restriction{\tilde J \equiv J + {L p^\chi \over 2\pi} \in  \IZ.}
 As a result, the support of the wavefunctions \wavefunctions\ at $\xp =0$
is $x\in  \IZ / p^+ - {L p^\chi / 2\pi p^+}$, i.e. the lattice has a $p^\chi$-dependent shift.
This allows one to construct a Fock space basis consisting only of wavefunctions
which are smooth everywhere. We will be introduce it in section 6.

%%%%%%%%%%%%%%%%%%%%%%%%%%%%%%%%%%%%%%%%%%%%%%%%%%%%
\newsec{ Strings on the smooth orbifold}
%%%%%%%%%%%%%%%%%%%%%%%%%%%%%%%%%%%%%%%%%%%%%%%%%%%%

The analysis of string propagation on the
orbifold $\ol$ proceeds
very similarly to the analysis for $\ozero$ performed in
\lms.

\subsec{Light-cone gauge quantization}

In the  light-cone gauge $x^+=y^+ = \tau$, we can express the worldsheet
Lagrangian for as
 \eqn\lclagrangian{
%%\eqalign{
 {\CL}= - p^+ \partial_\tau x^-_0 + {1\over 4\pi \alpha'}
 \int_0^{2\pi}
 d\sigma \left(\alpha' p^+
\left(\p_\tau x \p_\tau x + \p_\tau \chi \p_\tau \chi \right)
 - {1\over \alpha'p^+ }
\left(\p_\sigma x \p_\sigma x + \p_\sigma \chi \p_\sigma \chi \right)
\right) .}
%%\cr
%% & = -p^+ \partial_\tau y^-_0 + {1\over 4\pi \alpha'}
%% \int_0^{2\pi}
%% d\sigma \tau^2\left(\alpha' p^+ \p_\tau y\p_\tau y
%% - {1\over \alpha'p^+} \p_\sigma y \p_\sigma y \right)}}
%where
% \eqn\xyrelli{\eqalign{
% &x(\sigma,\tau)=\tau y(\sigma,\tau) \cr
% &y^-_0(\tau)=x^-_0(\tau)-{1 \over 2\tau} \int_0^{2\pi}
%  {d\sigma \over 2\pi}  (x(\sigma, \tau))^2}}
Invariance under the choice of origin of the $\sigma$-coordinate leads to the constraint
\longpaper
 \eqn\lcconstraint{\int d\sigma \left(
\p_\sigma \chi \p_\tau \chi +
\partial_\sigma x \partial_\tau  x
- {1\over 2\tau} \partial_\sigma x^2 \right) =0. }
The orbifold action
 \eqn\lcorbifold{\eqalign{
 &x(\sigma,\tau) \to x(\sigma,\tau) +2\pi n \tau\cr
 &x^-_0 (\tau) \to x^-_0(\tau) + 2\pi n \int_0^{2\pi}
 {d \sigma \over 2\pi}
 x(\sigma,\tau) +{(2\pi n)^2\over 2} \tau \cr
 & \chi(\sigma, \tau) \to \chi(\sigma, \tau) + L,}}
%\cr
% &y(\sigma,\tau)\to y(\sigma,\tau) + 2\pi n}}
leaves both \lclagrangian\ and \lcconstraint\ invariant.

Because of the invariance of the action \lclagrangian\
under constant shifts of $\xm$, $p^+$
is a conserved quantity. The equation of motion for $p^+$ implies
 \eqn\lchamiltonian{
%%\eqalign{
 P_{x^+}=p^+\partial_\tau x^-_0 = {1\over 4\pi\alpha'}
 \int_0^{\ell} d\sigma \left( (\p_\tau x)^2 + (\p_\tau \chi)^2
 + (\p_\sigma x)^2 +  (\p_\sigma \chi)^2 \right) }
%% \cr
%% &P_{y^-}=p^+\partial_\tau y^-_0 ={1\over 4\pi\alpha'}
%%  \int_0^{\ell} d\sigma \tau^2 \left(\p_\tau y\p_\tau y
%% + \p_\sigma y \p_\sigma y \right)}}
where we have rescaled $\sigma$ to range in $[0,\ell=2\pi
\alpha'p^+)$. The direction of the Killing vector
$\xp$ is changed by the orbifold action, and as a result,
the Hamiltonian $P_{\xp}$is not invariant under \lcorbifold.

The mode expansions for the physical worldsheet fields in the
$w$-twisted sector can be written as follows:
 \eqn\xsol{\eqalign{
& x(\sigma,\tau) =\xi + {p \over p^+}\tau + {2\pi w
  \sigma \tau\over \ell}+ \cr
 & \qquad i \left({\alpha'\over 2}\right)^\half \sum_{n\not= 0}
 \left\{
{\alpha_n \over n}
\exp \left[- {2\pi in(\sigma+\tau) \over \ell}\right] +
{\tilde \alpha_n \over n}
\exp\left[ {2\pi in(\sigma-\tau) \over \ell}\right]
 \right\}\cr
& \chi(\sigma, \tau) =
\chi_0 + { p^\chi \over p^+} \tau
+ { w \sigma L \over \ell} + \cr
 & \qquad i \left({ \alpha' \over 2}\right)^\half
    \sum_{n \not= 0} \left\{
{a_n \over n}
\exp\left[ -{2 \pi in (\sigma+\tau) \over \ell }\right] +
{\tilde a_n \over  n}
\exp\left[ {2\pi in(\sigma-\tau) \over \ell}\right]
\right\},}}
where, upon quantization, the oscillators satisfy the usual commutation relations.
The solution of the equation of motion for $x^-_0$
is $ x^-_0 = P_{\xp} \tau / p^+$, up to an additive constant. As in the previous section,
$\tilde J$ defined as
\eqn\restrict{\tilde J \equiv J + {L \pc \over 2\pi} = -\xi p^+ + {L \pc \over 2\pi}}
must be quantized. Here, the constraint $\lcconstraint$ implies
\eqn\levelmatching{\tilde J w = N - \tilde N,}
where $N$ and $\tilde N$ are the usual number operators.

Strings on the $\ol$ orbifold can also be quantized covariantly
in a manner very similar to the covariant quantization of $\ozero$
in \lms.

\subsec{The torus partition function}

The calculation of the torus partition function for bosonic strings on
the resolved orbifold
${\IR^{1,3} / \Gamma_L}$ was actually implicitly contained in \lms.
It is, however, instructive to see it more explicitly.
We wish to calculate
\eqn\zdefinition{Z= \int [{\bf dX}] [d \chi]e^{ iS}}
with
\eqn\action{S ={1 \over 4\pi\ap} \int   d^2\sigma \sqrt{g}g^{ab}
\left( \p_a {\bf X}^T(\sigma_1, \sigma_2) \cdot {\bf G} \cdot \p_b {\bf X}(\sigma_1, \sigma_2)
 + \p_a  \chi(\sigma_1, \sigma_2)  \p_b  \chi(\sigma_1, \sigma_2)    \right).  }
Here the coordinates $x^+, x$ and $x^-$ are  combined into a single vector ${\bf X}$,
and \bG\ represents the corresponding part of the space-time metric:
\eqn\xvector{{\bX} = \pmatrix{ x^+ \cr x^{~}\cr x^-\cr}, \qquad \bG=\pmatrix{0&0&-1\cr
 0&1&0\cr-1&0&0}.}
The orbifold action becomes
\eqn\actiononX{\bX \rightarrow e^{2\pi \CJ} \bX = (1+2\pi \CJ + 2\pi^2 \CJ^2) \bX, \qquad
\CJ=\pmatrix{0&0&0\cr
 1&0&0\cr0&1&0}.}
The worldsheet metric can be chosen as
 \eqn\torusm{ g_{ab}\ d\sigma^a d\sigma^b= (d\si + \tau_+ d \sii)(d\si +
 \tau_- d\sii), \quad \tau_{\pm}= \tau_1 \pm \tau_2,}
with $\si,\sii \in [0,1)$. Plugging the world-sheet metric into \action,
we get
\eqn\actiontwo{\eqalign{S =  {1 \over 4\pi\ap \tau_2}\int d^2\sigma
( & \tau_+ \tau_- \p_1 {\bf X}^T \cdot {\bf G} \cdot \p_1 {\bf X} -
2 \tau_1 \p_1 {\bf X}^T \cdot {\bf G} \cdot \p_2 {\bf X} +
\p_2 {\bf X}^T \cdot {\bf G} \cdot \p_2 {\bf X} \cr
& + (\p_1  \chi)^2 + (\p_2  \chi)^2   ). \cr  }}
The worldsheet fields in the $(w_1, w_2)$-twisted sector can be expanded as
\eqn\torusfields{\eqalign{
\bX(\si,\sii) &= \exp\bigl[ 2\pi(\si w_1 + \sii w_2) \CJ \bigr]
\sum_{n_1, n_2 \in \IZ}
 \bX_{n_1, n_2} e^{2\pi i (n_1\si + n_2 \sii)} \cr
\chi(\si,\sii) &= \si w_1 L  + \sii L \tau_2^{-1}(w_2 - w_1 \tau_1 ) +
\sum_{n_1, n_2 \in \IZ}
 \chi_{n_1, n_2} e^{2\pi i (n_1\si + n_2 \sii) },\cr }}
implying, in particular,
\eqn\torusderivatives{\eqalign{
\p_1\bX(\si,\sii) &= 2\pi w_1 \CJ \bX + \exp\bigl[ 2\pi(\si w_1 + \sii w_2) \CJ \bigr]
\sum_{n_1, n_2 \in \IZ}
 2\pi i n_1 \bX_{n_1, n_2} e^{2\pi i (n_1\si + n_2 \sii)} \cr
\p_2\bX(\si,\sii) &= 2\pi w_2 \CJ \bX + \exp\bigl[ 2\pi(\si w_1 + \sii w_2) \CJ \bigr]
\sum_{n_1, n_2 \in \IZ}
 2\pi i n_2 \bX_{n_1, n_2} e^{2\pi i (n_1\si + n_2 \sii)}. \cr }}

Let us first see how the modes of $\bX$ with non-zero
$(n_1,n_2)$
contribute to the partition function. Their contributions
to the action will be of the form
\eqn\qfluctuations{\sum_{\alpha, \beta = 0,1,2}
c_{\alpha \beta}\ \bX^T_{-n_1,-n_2} \cdot (\CJ^T)^\alpha \cdot \bG \cdot \CJ^\beta \cdot \bX_{n_1,n_2}.   }
All the matrices $ (\CJ^T)^\alpha \cdot \bG \cdot \CJ^\beta$ are
of the form
$$ \left( \matrix{  \Upsilon_{11} & \Upsilon_{12} & \Upsilon_{13}\cr
         \Upsilon_{21} & \Upsilon_{22} & 0 \cr
         \Upsilon_{31} & 0 & 0 }  \right),$$
and if $\alpha$ or $\beta$ is non-zero, then $\Upsilon_{13} = \Upsilon_{22} = \Upsilon_{31} =  0.$
Therefore only  terms with
$\alpha =\beta=0$
will contribute to the determinant of the matrix and have any
effect on the result of the gaussian integration\foot{We wick rotate the timelike modes to define this integral.}.
Because they are independent of $w_1$ and $w_2$, one concludes that the contribution of
oscillators of $\bX$ to the partition function will be the same as in Minkowski space \lms.

As a result, only the zero modes of $\bX$ can have any $(w_1,
w_2)$-dependent contribution. Because in this case $(n_1, n_2 ) = (0,0)$, the second terms on
the RHS of \torusderivatives\ vanish, and we get
\eqn\actionthree{S_{X_{0,0}} =  {(2\pi)^2 \over 4\pi\ap \tau_2} \int d^2\sigma
(  \tau_+ \tau_-  w_1^2   -
2 \tau_1  w_1 w_2   +
 w_2^2 ) \bX^T_{0,0} \CJ^T  {\bf G}   \CJ \bX_{0,0} ,   }
or\foot{To avoid any confusion, we should stress that here $\xp$
denotes the zero-mode of $\xp(\sigma, \tau).$}
\eqn\actionfour{S_{X_{0,0}} =  {\pi \over \ap \tau_2}
(  w_1 \tau_+ -w_2) (w_1 \tau_- - w_2) (\xp)^2.}
Since in each sector $\chi$ is just a periodic boson, whose zero mode
contributes an action
\eqn\chiaction{S_{\chi_{0,0}} =  {\pi \over \ap \tau_2}
(  w_1 \tau_+ -w_2) (w_1 \tau_- - w_2) {L^2\over 4\pi^2},}
we find that the full partition function for the bosons
is
 \eqn\partitionfunction{ \eqalign{ Z =
 &  {iZ^{\rm ghost}Z^{\perp }L \over
 \tau_2^{2} (\eta(\tau_+)\eta(-\tau_-))^{4}} \ \times \cr
& \int ~~ \! \! \!
 {d^3 x \over (2\pi\sqrt{\ap})^4} \! \! \! \sum_{w_1,w_2\in \IZ} \!\! \! \exp\Biggl[{ i \pi (  w_1 \tau_+ -w_2) (w_1 \tau_- + w_2)\over \ap \tau_2}
 [(\xp)^2 + {L^2\over 4\pi^2}] \Biggr].
 \cr}
}

 Using Poisson resummation, one can show \lms\ that for $L=0$ the bosonic
 contribution to the one-loop cosmological ``constant''
 diverges as $\xp \to 0$,
\eqn\divergence{
\Lambda(x^+)\sim {1\over (x^+)^2 } \times
\int_{{\cal F}} {d^2 \tau  \over (\tau_2)^2}  {Z^{\rm tr}\over
 \vert \eta(\tau)\vert^2} }
For $L \neq 0$, this contribution is resolved to
\eqn\nodivergence{
\Lambda(x^+)\sim {1\over (x^+)^2 + {L^2 \over 4\pi^2}} \times
\int_{{\cal F}} {d^2 \tau  \over (\tau_2)^2}  {Z^{\rm tr}\over
 \vert \eta(\tau)\vert^2} + \dots,   }
where the dots denote other non-singular terms (which are
subleading if $L$ is small).
This is consistent with the interpretation \lms\ of
the divergence at $L = 0, x^+ = 0$ as arising
from light winding modes.

\subsec{Extension to superstring backgrounds}

The $\ol$ orbifold can be thought of as a superstring background.
In the Green-Schwarz formalism
one needs to add six more worldsheet bosons (corresponding to $\IR^6$)
and the appropriate worldsheet
fermion content.
Because $\pi_1(\ol) = \IZ$, there are two different spin structures we
can choose. One of them makes the orbifold supersymmetric, and corresponds to
all the extra worldsheet fields being single-valued \longpaper.
For the non-supersymmetric choice, strings with odd winding
numbers will have antiperiodic worldsheet fermions.

The formulas for the bosonic partition function \partitionfunction\
can be applied to the superstring. The full partition function
will contain  a factor of the fermionic partition function, which
will vanish if we choose the supersymmetric spin structure.

With the opposite choice of the spin structure, supersymmetry is
broken and a one-loop dilaton tadpole is generated, in analogy to
\rohm.  For large enough $L$, this potential is an
everywhere-finite function on spacetime which goes to zero at
large $|x^+|$. However, because of the low codimension of its
support, this one-loop energy-momentum tensor causes a significant
backreaction on the geometry.
To avoid this, one could consider including a rotation of the
extra planes ($x^4$-$x^5$, $x^6$-$x^7$, $x^8$-$x^9$) into the
action of the orbifold group generator ${\rm g}_L$. Compensating
for the Casimir stress-energy by a small local change of the
metric and dilaton near the axes of rotation would be similar to
\aharony.

In this way, one would obtain a non-supersymmetric perturbatively
stable time-dependent background. At least in some range of
parameters, the leading instability would be nucleation of Witten
bubbles \refs{\bubbles \dggh \DowkerGB \DowkerGB \ShinkaiAS
\FabingerJD \HorowitzGN \DeAlwisKP \BirminghamST -
\BalasubramanianAM, \aharony}. The fact that such bubbles are
possible (in some range of parameters) can be seen quite easily.
Given a small enough $v$, the region around $\tilde x = \tilde x^+
= 0$ looks like a time-independent generalized `twisted circle'
orbifold \refs{e.g.\ \TseytlinEI \RussoIK \CostaNW \GutperleMB
\RussoTF \MotlDJ \TseytlinQB - \DavidVM} (cf. \tmetric), the twist
being the rotation in the extra planes ($x^4$-$x^5$, $x^6$-$x^7$,
$x^8$-$x^9$). As a result, it must be possible to nucleate a
generalized Witten bubble locally. For this reason, there must
exist a bubble solution which is extendable  to a global solution
to Einstein's equations asymptoting to the non-supersymmetric
time-dependent orbifold. It is possible that such a solution could
be obtained by an alternative Wick rotation of the Euclidean Kerr
solution.

%%%%%%%%%%%%%%%%%%%%%%%%%%%%%%%%%%%%%%%%%%%%%%%%%%%%
\newsec{Tree-level string interactions
on the smooth orbifold}
%%%%%%%%%%%%%%%%%%%%%%%%%%%%%%%%%%%%%%%%%%%%%%%%%%%%

The basis of states used in \lms\ to calculate string
amplitudes on the $\ozero$ orbifold is a singular one.
Using this basis of states
in Minkowski space one would find similar tree-level
divergences in special kinematic regimes.
However, on the singular orbifold one has no option
(as one has in Minkowski space)
to use a less singular basis.

On the smooth orbifold, one has a better option.
Define a basis of wavefunctions by the following
convolution of plane waves:
\eqn\goodbasis{\eqalign{ \psi_{\chi_0, \tilde{J}, p^+, \vec{p}}&
(x^+, x, x^-, \chi, \vec{x}) \equiv \cr & \sqrt{a \over 2 \pi}
\int d \pc e^{- i \pc \chi_0 - {a \over 4} (\pc)^2 } \int {dp
\over \sqrt{2 \pi}} ~~e^{ - i p \xi} \phi_{p^+, p, \pc} ~(x^+, x,
x^-, \chi) e^{i\vec{p}\cdot\vec{x} },\cr}}
where in this integral, the variable $\xi = - J/p^+$ is considered
as a function of $\pc$ and $\tilde{J}$, via the equation
$$ \xi \equiv {1\over p^+ }( - \tilde{J} + {L \over 2 \pi} \pc  ) ;$$
and $\phi_{p^+, p, \pc} ~(x^+, x, x^-, \chi)$ is an on-shell plane
wave as in \phidef, with $p^-$ now including also a contribution
from $\vec{p}$.

Performing the two integrals in \goodbasis\ one finds
\eqn\ilovedoinggaussianintegrals{ \eqalign{ \psi_{\chi_0,
\tilde{J}, p^+, \vec{p}}& (x^+, x,  x^-, \chi, \vec{x}) = e^{ i
p^+ x^- + i {\xp \over 2 p^+} ( \vec{p}^2 + m^2) + i \vec{p} \cdot
\vec{x} + {i p^+ \over 2 \xp} \left( x + {\tilde{J} / p^+ }
\right)^2 } ~~\times \cr &\times{\sqrt{a} |p^+| \over \sqrt{ 2
\left(\lslash\right)^2 - i  a x^+ p^+ - 2 (x^+)^2 }} ~~\times \cr
&\times \exp \left[ {- \xp p^+  \over -2 i\left(\lslash\right)^2 +
\xp p^+ a + 2i (\xp)^2}
%\times \right.  \cr
%\!\!\!\!\!& \left.
\left(
\chi - \chi_0 - {L \over 2 \pi \xp} \left( x + {\tilde{J} / p^+} \right)
\right)^2 \right]
%- ( \chi - \chi_0)^2 +
%2 (\chi - \chi_0) \lslash ( {\tilde{J} \over p^+ \xp } - 1)
%- \left(\lslash\right)^2
%\left(1 - { \tilde{J} \over p^+ \xp } \right)^2 \right) \right]
}}
We note that this convolution (which diagonalizes $\tilde{J}$)
would not be sensible on the orbifold without the shift by $L$. As
one can see from the fact that the magnitude of this wave remains
finite near $\xp \to 0$, these wavefunctions lack the focusing
properties of those in the singular orbifold. (Intuitively, they
are gaussian linear superpositions of $J$-eigenfunctions
\wavefunctions, \psili\ which are focused on different points.)

As one might expect from this observation, the tree-level
divergences in string theory amplitudes involving untwisted states
on the singular orbifold are absent in the resolved case using
this basis of states. For example for tachyons the four-point
amplitude for scattering these  states is obtained, as in \lms, by
convolving the four-point amplitude for plane-wave tachyons on the
covering space with the kernel which produces the wavefunctions
from plane waves, and replacing delta functions with Kronecker
deltas. The amplitudes in the $\tilde J$ basis
\ilovedoinggaussianintegrals\ may be obtained from those in the
$J$-basis (equation (6.16) of \lms) by a  convolution integral
over $\pc$ considering the quantum number $J =  \tilde J - \lslash
\pc$ to be a function of $\pc$.

Amplitudes involving generic momenta are smooth \lms.
A divergence was found in a kinematic regime
which explored the Regge region of the amplitude
on the covering space.
The dangerous part of the amplitude occurs when
$p_1^+ = p_3^+$, near very large $q$ in the following integral:
\eqn\fourpointone{
\CA_4 \sim \int^\infty {dq \over |q|}
\int dk
~~e^{\half i k \chi^t - {a \over 8} k^2 }
~~q^{ 4 - \ap (k^2 + \vec{p}_t^2 )}
~~\exp\left( - i q \sqrt{\mu_{12}} \xi_t \right)
%\exp\left( - i q {\sqrt \mu_{12} \over p_1^+} (- \tilde{J}_t + \lslash k) \right)}
}
where $k \equiv \pc_3 - \pc_1$,
$\tilde{J}_t \equiv \tilde{J}_3 - \tilde{J}_1$, $\chi^t = \chi_0^1 - \chi_0^3$  and
\eqn\xit{\xi_t
\equiv \xi_3 - \xi_1 = {1 \over p_1^+} ( - \tilde{J}_t + \lslash k)
}
Performing the integral over $k$ one finds
\eqn\nodivergence{
\CA_4 \sim \int^\infty {dq \over |q|}
{1 \over \sqrt{a + 8 \ap \ln q} }
\exp\left[
- {2 ( {\sqrt{\mu_{12}} \over p_1^+ } \lslash q - \half \chi^t )^2  \over
    a + 8 \ap \ln q }
+ \ln q ( 4 - \ap \vec{p}_t^2 ) +
i {\sqrt{ \mu_{12}} \over p_1^+ } \tilde{J}_t q \right]}
which is well-behaved near $q \to \infty$ because of the
$- L^2 q^2 / \ln q$ term
in the exponential.

The divergences in four-point functions found in \lms\
are therefore absent in the resolved orbifold.
One might have worried that these divergences were
associated purely with the time-dependent nature of the
background
%i.e. that in time-dependent backgrounds
%one is forced to work with some basis of wavefunctions
%for which some amplitudes are singular.
and the identifications which include very large boosts.
This calculation
demonstrates that this is not the case.

\newsec{On the nature of the resolution by the shift}

The light-like orbifold singularity of \lms\ is very different from
any time-like singularity in string theory, such as the $A_k$ singular limit
of the supersymmetric $\IC^2 / \IZ_{k+1}$ orbifold.
In particular, if one wants to obtain the $A_k$ singularity
as a limit of some smooth geometry, one has to send the curvature to infinity. In this sense
the $A_k$ singularity is really a curvature singularity. This is not true in the case of
 the $\IR^{1,3} / \Gamma_0$ orbifold, because it can be thought of as the $L \rightarrow 0$ limit
 of the $\IR^{1,3} / \Gamma_L$ orbifold, which is smooth and flat everywhere.

It is important that the group $\Gamma_0 = \IZ$ \orbifoldaction\ is not finite.
If one added a shift by $L$ in an extra direction, for example, to the orbifold group generator of $\IC^2 / \IZ_{k+1}$,
the  group would become $\IZ$ and one would obtain a (generalized) `twisted circle' orbifold
\refs{\TseytlinEI \RussoIK \CostaNW \GutperleMB \RussoTF \MotlDJ
\TseytlinQB - \DavidVM}. In that case,
sending $L$ to
zero would not lead to the  original $\IC^2 / \IZ_{k+1}$. Points identified by
the $(k+1)$-th power of the orbifold group generator would be very close to each other,
and for any non-zero $L$, they would be distinct.
If we wanted the get some $\IC^2 / \IZ_{k+1}$ orbifold in the limit $L \rightarrow 0$, we would be
forced to perform a T-duality.

\subsec{On the possibility of a local resolution by the shift mode}

Because the $\ozero$ orbifold is a limit of the smooth $\ol$ orbifold, it is natural to ask whether
there are any physical modes in the $\ozero$ orbifold which would make $L$ locally non-zero
and smooth out the null singularity.

To address this question we calculate the distance in the configuration space between  $L = 0$ and some fixed non-zero
$L = L_f$. More precisely, we will calculate the Zamolodchikov metric for the CFT operator
which increases $L$ by $\delta L$
in the light-cone gauge at some fixed non-zero light-cone time, and then integrate it over $L$.
In the time-independent context,
a similar calculation for a marginal operator would give the distance in the moduli space.

Let us first use the $y$ and $\psi$ coordinates \ycoordinates, \ymetric.
The worldsheet action (with the period of $\sigma$ being $2\pi\ap p^+$)
\eqn\wsaction{S =\!  {1 \over 4\pi \ap} \!\int \! \! \! d^2\sigma \left(
(\tau^2 + {L^2\over 4\pi^2})[(\del_\tau y)^2 - (\del_\sigma y)^2)]
+ [(\del_\tau \psi)^2 - (\del_\sigma \psi )^2] + {L\over \pi} [ \del_\tau y  \del_\tau \psi  - \del_\sigma y \del_\sigma\psi]  \right),   }
leads to the correlation functions
\eqn\correlators{\eqalign{
\langle y(\sigma, \tau) y(\sigma', \tau')\rangle &= -{\ap \over 2\tau \tau'} \ln |(\tau - \tau')^2 - (\sigma - \sigma')^2| \cr
\langle y(\sigma, \tau) \psi(\sigma', \tau')\rangle &= {\ap L \over 4 \pi \tau \tau'} \ln |(\tau - \tau')^2 - (\sigma - \sigma')^2| \cr
\langle \psi(\sigma, \tau) \psi(\sigma', \tau')\rangle &=
-{\ap \over 2} \left(1 +  {L^2 \over 4\pi^2\tau \tau'}\right) \ln |(\tau - \tau')^2 - (\sigma - \sigma')^2| .\cr }}
The operator which, when added to the worldsheet action, changes $L$ by $\delta L$ is
\eqn\operator{\CO_L\delta L = {1 \over 4 \pi \ap }
\left({L\over 2\pi^2 } ( (\del_\tau y)^2 - (\del_\sigma y)^2) +
{1 \over  \pi}
( \ \del_\tau y  \ \del_\tau \psi  -  \del_\sigma y \ \del_\sigma\psi)
\right)  \delta L.}
Using \correlators, one finds that the Zamolodchikov metric, defined as
\eqn\zamolodchikov{\CG _{L,L}\ \delta L \ \delta L \equiv
 \lim_{\epsilon \to 0} \epsilon^4 \langle \CO _L(\sigma, \tau) \CO _L(\sigma +  \epsilon, \tau) \rangle \ (\delta L)^2,
  }
is
\eqn\zamoltwo{\CG _{L,L}\ \delta L \ \delta L =
 {1 \over  2 \pi^2 (y^+)^2}  \ (\delta L)^2,
  }
The resulting distance between $L=0$ and $L = L_f$ is
therefore
\eqn\distance{\Delta \CD={ L_f \over
\sqrt{2} \pi
|\yp| }. }

An easy way to arrive at \zamoltwo\ is
to translate the Zamolodchikov metric
\zamolodchikov\ into the
usual metric on the space of metrics \candelasetal.
The marginal operator \operator\ can be written in a
general form
\eqn\operatortwo{\CO_L\delta L ={1\over 4\pi \ap}\delta G_{\mu \nu}\ (\p_\tau Y^{\mu} \p_\tau Y^{\nu} - \p_\sigma Y^{\mu} \p_\sigma Y^{\nu}), }
where $\mu, \nu = 2,3$ and where we have defined $Y^2 = y$ and $Y^3 = \psi$. The coordinates
$y$ and $\psi$, defined in \ycoordinates, are
linear combinations of the cartesian coordinates $x$ and $\chi$ whose coefficients
depend on $\tau = \xp$. We can express this fact as
\eqn\yandx{Y^{\mu}(\sigma, \tau) = e^{\mu}_{\alpha} (\tau) X^{\alpha}(\sigma, \tau),}
where $\alpha = 2,3$ and $X^2 = x, X^3 = \chi$. This allows us to rewrite
the expectation value of interest as
\eqn\expectation{\eqalign{ \langle \CO _L(\sigma, \tau)&  \CO _L(\sigma +  \epsilon, \tau) \rangle \ (\delta L)^2
 = {1 \over (4\pi \ap)^2 }\delta G_{\mu \nu} \ e^{\mu}_{\alpha} (\tau) e^{\nu}_{\beta}(\tau)
  \ \delta G_{ \mu^\prime  \nu^\prime}\ e^{\mu^\prime}_{\alpha^\prime}(\tau) e^{\nu^\prime}_{\beta^\prime}(\tau)\ \times
   \cr
& \times \ \langle [\p_\tau X^{\alpha} \p_\tau X^{\beta} - \p_\sigma X^{\alpha} \p_\sigma X^{\beta}] |_{\sigma, \tau}
 [\p_\tau X^{\alpha^\prime} \p_\tau X^{\beta^\prime} - \p_\sigma X^{\alpha^\prime} \p_\sigma X^{\beta^\prime}] |_{\sigma + \epsilon, \tau} \rangle +
   \dots }}
The dots here denote terms  containing $\tau$-derivatives of various $e^{\mu}_{\alpha}$. It is important
to note that such terms do not contribute to the limit \zamolodchikov\ and can be ignored.
Evaluation of the expectation values on the right-hand side of \expectation\
is identical to the standard calculation in flat Minkowski space, giving
\eqn\minkcorrelators{
\eqalign{
\langle [\p_\tau X^{\alpha} \p_\tau X^{\beta} - \p_\sigma X^{\alpha} \p_\sigma X^{\beta}] |_{\sigma, \tau}
 [\p_\tau X^{\alpha^\prime} \p_\tau X^{\beta^\prime} -
\p_\sigma X^{\alpha^\prime} \p_\sigma X^{\beta^\prime}] |_{\sigma + \epsilon, \tau} \rangle  \cr
= {2  \ap^2 \over \epsilon^4} (\eta^{\alpha \alpha^\prime } \eta^{\beta\beta^\prime}
+ \eta^{\alpha \beta^\prime} \eta^{\alpha^\prime \beta}). }}

The  Zamolodchikov metric then becomes
\eqn\zamolthree{\CG _{L,L}\ \delta L \ \delta L = {1\over 4 \pi^2} \ \delta G_{\mu \nu} \
e^{\mu}_{\alpha} \eta^{\alpha \alpha^\prime}e^{\mu^\prime}_{\alpha^\prime}
  \ \delta G_{ \mu^\prime  \nu^\prime}\ e^{\mu^\prime}_{\beta}\eta^{\beta \beta^\prime}e^{\nu^\prime}_{\beta^\prime},}
or
\eqn\zamolfour{\CG _{L,L}\ \delta L \ \delta L = {1\over 4 \pi^2} \ \delta G_{\nu \mu} \
G^{\mu \mu^\prime}
  \ \delta G_{ \mu^\prime  \nu^\prime}\ G^{\nu^\prime \nu}.}
Plugging
\eqn\metricg{G_{\mu \nu} =
\pmatrix{\tau^2 + {L^2 \over 4\pi^2} &  {L \over 2\pi} \cr {L \over 2\pi} & 1 \cr}, \ G^{\mu \nu}  = {1\over \tau^2}
 \pmatrix{1 &  -{L \over 2\pi} \cr -  {L \over 2\pi} & {L^2 \over 4\pi^2}  \cr}, \
\delta G_{\mu \nu} =\pmatrix{ {L\over 2\pi^2} & {1 \over 2\pi} \cr  {1 \over 2\pi} & 0 \cr } \delta L}
into \zamolfour, we get,
\eqn\zamolfive{\CG _{L,L}\ \delta L \ \delta L = { (\delta L)^2  \over  2 \pi^2 \tau^2}. }
Upon translating from light-cone gauge, $y^+ = \tau$, this gives \zamoltwo.

The reader may be puzzled that even in the completely smooth orbifold $\ol$ we have found a quantity
which diverges near $y^+ = 0 $.  This is an artifact of our choice of coordinates on field space.
If we chose instead
to modify $L$ using the coordinates $\xp, x, \xm$ \coord, \metric, and $\chit = \chi / L$,
we would write
\eqn\tildedeltaL{
\tilde \CO_L \delta L = { 1  \over 2 \pi \ap} [(\p _\tau \chit)^2 - (\p _\sigma \chit)^2]L\ \delta L. }
For the metric on field space we find (just like in the case
of changing the size of an ordinary Kaluza-Klein circle in string theory),
\eqn\zmetrictilde{
\tilde \CG_{L, L} \delta L \delta L \sim { (\delta L)^2 \over L^2 } .
}

Therefore, for any nonzero $L$ {\it or } nonzero $x^+$, we have found a fluctuating
mode which turns on the shift.  However, when both $L = 0$ and $ x^+ = 0$, the kinetic
term for any such mode diverges, and the fluctuations freeze out.
Further, as with any physical description which uses constant-value surfaces of
a null coordinate $x^+$ as initial-data slices,
objects with $p^+=0$ are subtle.  The singularity
at $x^+ = 0$ is such an object.
Begin with any spacelike hypersurface on which
to specify initial data with $L=0$.  If we try to extend it
as much as possible, it will
inevitably  touch the locus $x^+=0$.  At $\xp=0$
the $L$-mode which resolves the
orbifold does not fluctuate.  Therefore with this initial data,
$L$ will continue to vanish
along $x^+=0$,
 and the singularity will persist.

We should point out that our discussion above is not completely conclusive.
The most important issue is whether for $L=0$ we are allowed to
do any perturbative calculations  near the singularity at all.
It is natural to expect that in the near singularity region any string probe
would cause a large backreaction and make the singularity locally
space-like, in the spirit of \LawrenceAJ. For this reason, the applicability
of CFT techniques is questionable if $L$ and $|\xp|$ are both small.

\newsec{Replacing the $\ozero$ singularity with a sandwich wave}

In this section, we would like to ask whether it is possible to have a smooth future evolution of the
geometry if we specify the initial data on the slice of a constant light-cone time $\yp$ to be exactly those of
the (singular) $\ozero$ orbifold. In other words, is the presence of the singularity inevitable once
such initial data on $y^+ = y^+_{in} < 0$ are given?

As we will see, the answer is that the singularity at $\yp = 0$
can be evaded easily and we will construct the corresponding
gravitational solutions
%%b
(which are actually exact to all orders in $\ap$ \HorowitzBV).
%%e
 The $y^+ =
y^+_{in}$ hypersurface is not a Cauchy surface, and of course,
there can be gravitational waves coming from the region where the
data have not been specified.

The appropriate classical gravitational solutions are smooth orbifolds of a certain type
of four-dimensional Ricci-flat ``sandwich waves'' \sandwich\ described in \rindler\ (times
$\IR^6$). These waves belong to the family of plane-fronted waves with parallel rays (pp-waves), as can be seen from \ppwaves.
They consist of three regions: (I) $\yp < A$, where the spacetime is exactly flat, (II)
$\yp \in (A,B)$ with the Riemann tensor non-vanishing, and  (III) $\yp > B$, where the spacetime is flat again. The support
of the Riemann tensor is therefore perfectly localized between $A$ and $B$.

We will parametrize the geometry in all the three regions by one set of coordinates.
In addition to  $\yp$, we will  also use coordinates $\ym, y$ and $ \chih$.
The metric in  region II can be written as
\eqn\metricII{ds^2_{II}= -2 d\yp d\ym + N^2_y \cosh^2[\Omega(\yp - M)]  dy^2  + N^2_\chih \cos^2[\Omega(\yp - M)] d\chih^2, }
where $M, N_y, N_\chih$ and $\Omega$ are some constants. This metric can be matched onto
flat Minkowski space along $\yp = A$ and $\yp = B$. In order to do so, one has
to perform a coordinate change to obtain the following
form of the Minkowski metric in regions I and III
\eqn\metricIandIII{\eqalign{ds^2_{I} &= - 2 d\yp d\ym + (\alpha_I +  \yp\beta_I)^2 dy^2 + (\gamma_I +  \yp\delta_I)^2 d\chih^2 \cr
ds^2_{III} &= - 2 d\yp d\ym + (\alpha_{III} +  \yp\beta_{III})^2 dy^2 + (\gamma_{III} +  \yp\delta_{III})^2 d\chih^2.}}
The appropriate coordinate transformation relating \metricIandIII\ and
\eqn\flatspace{ds^2 = - 2dUdV + (d\tilde X^2)^2 + (d\tilde X^3)^2}
is
\eqn\coordinatechange{\eqalign{U &= \yp \cr V &= \ym + \half \beta (\alpha +  \yp\beta) y^2 + \half \delta (\gamma +  \yp\delta) \chih^2 \cr
 \tilde X^2 &= (\alpha +  \yp\beta)y \cr \tilde  X^3 &= (\gamma +  \yp\delta) \chih}}
The matching conditions requiring that the metric and its first derivatives are smooth  lead to
 certain simple algebraic relations between the coefficients in \metricII\ and \metricIandIII,
which we will not write down explicitly.
In addition, there is a constraint resulting from the requirement that the metric have no coordinate singularity
in region II.

Intuitively, the distance between two points separated by a constant coordinate distance in $y$ changes linearly
with $\yp$ in region I, then it follows a hyperbolic cosine in region II, and finally in region III, it changes linearly again.
(See fig. 1.) %%%%%%%%%%%%%%%%%%   \munch
A similar statement is true for the $\chih$ coordinate, with the cosine not being hyperbolic.

Note that the metric  \metricII, \metricIandIII\ in all the three regions is invariant under constant
shifts of $y$. This allows us to  consider a $\IZ$ orbifold with the orbifold group action generated by
$y \to y + 2\pi$,
\eqn\identif{(\yp, y, \ym, \chih) \sim (\yp, y + 2 \pi, \ym,  \chih).}
%
% (Of course, the same is true for $\chi$. We will, however, compactify only $y$.)
In particular
for region I, we may choose $\alpha_I = 0$ and $\beta_I  = 1$:
\eqn\metricI{ds^2_{I} = - 2 d\yp d\ym + (\yp)^2 dy^2 + (\gamma_I +  \yp\delta_I)^2 d\chih^2.}
Using a coordinate transformation in the spirit of \coordinatechange, we can write this also as
\eqn\metricIagain{ds^2_{I} = - 2 d\yp d\ym + (\yp)^2 dy^2 +  d\chi^2.}
In this way, we obtain exactly the same metric and identifications as in the case of the $\ozero$ orbifold, expressed in terms
of coordinates $\yp, y, \ym, \chi$ defined by \coord\ and \ycoordinates\ with $L=0$.

If we start with \metricIagain\ (or equivalently with \metricI),
what will be the geometry after the wave passes, i.e. in region III?
This depends on the strength of the
wave we choose and on the light-cone time $B$ where
the wave is glued
to flat space.

If we choose $B-A$ small, the $y$-circle will continue to shrink after the wave passes (slower than before),
 and eventually it will form a singularity.
On the other hand, an appropriate choice of parameters ($B>M$) can make the $y$-circle
expand again without going through a singularity. This can be seen as follows.

\ifig\munch{
The circumference of the $y$-circle is proportional to $\sqrt{g_{yy}}$.
In the non-singular case it first linearly contracts just like in
the $\ozero$ orbifold, then it follows a hyperbolic cosine, and eventually,
it linearly expands (or stays constant, in the marginal case).}
{\epsfxsize3.0in\epsfbox{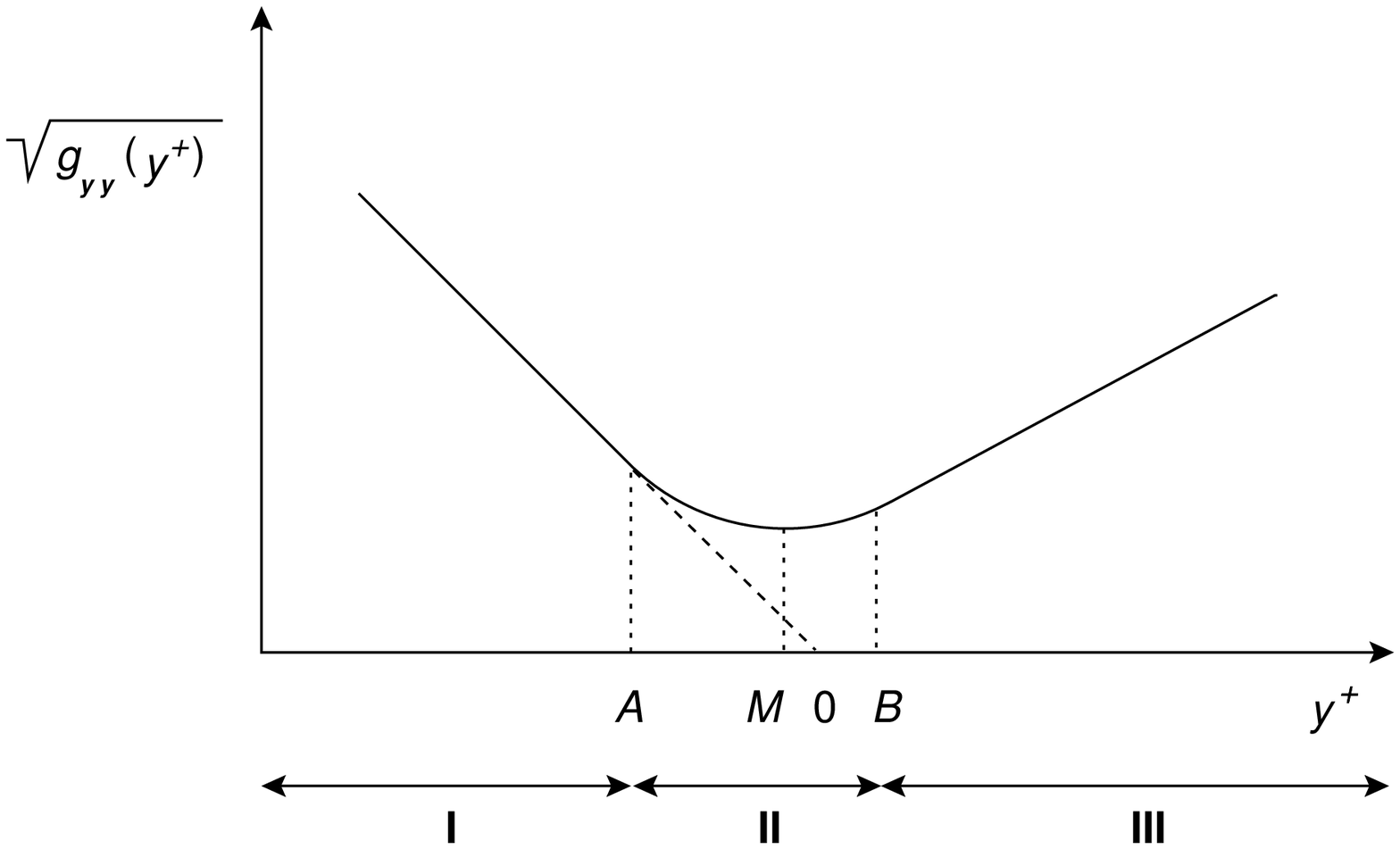}}

The matching conditions at $\yp = A$ ($A<0$) imply
\eqn\matching{
\Omega A \tanh [\Omega (A-M)] = 1, \quad N^2_y = {A^2 \over \cosh^2[\Omega( A- M)].}
}
This means that if we choose some definite value of $\Omega >1 / |A|$, the value of $A-M$
 and $N^2_y$ will be uniquely determined. In addition, in order to avoid $\chih$-coordinate singularities
 between $\yp= A$ and $\yp = M$, we need to satisfy
\eqn\nonsingular{ \Omega |A-M| < {\pi \over 2}, } which using
\matching\ is equivalent to \eqn\nonsingularII{ \Omega >   {1
\over \tanh({\pi \over 2})|A|}. } The condition that the null
singularity is removed by the sandwich wave is that function
$\sqrt{g_{yy}^{III}} = |\alpha_{III} + y^+ \beta_{III}|$ not
vanish inside region III. If we make the natural choice
$\alpha_{III} + \beta_{III}B
>0$, this implies that $\beta_{III} \ge 0$. By the matching
condition at $y^+ = B$ (analogous to a combination of equations
\matching)
$$
N_y^2\ \sinh[\Omega(B-M)]\ \cosh[\Omega(B-M)] = (\alpha_{III} + \beta_{III}B)\ \beta_{III},
$$
the requirement $\beta_{III} \ge 0$ is equivalent to $B \ge M$. Of
course, this can also be seen very intuitively from \munch.

We see that if we choose $\Omega$ consistent with \nonsingular,
determine $M$ and $N^2_y$ from \matching, and take $B > M$ (and
$\Omega (B-M) < \pi /2$ to avoid $\chih$-coordinate singularities
inside $\yp \in [M,B]$),
 we obtain a wave where the circumference of the $y$-circle first shrinks and then it expands without ever vanishing.
 For the marginal choice $B = M$, the
light-cone time future of the geometry will be a static Kaluza-Klein compactification on a circle.

%%%%%%%%%%%%%%%%%%%%
\newsec{Conclusion}
%%%%%%%%%%%%%%%%%%%%

In this paper, we have studied a completely smooth time-dependent orbifold
of Minkowski space.  Strings in this background are very
well-behaved, and the string amplitudes appear to define
a set of S-matrix observables.
In particular, perturbative corrections to the background are
well-controlled by the value of the dilaton at large $|\xp|$.
This is true even in certain nonsupersymmetric generalizations, with
large enough $L \gg \sqrt{\ap}$.
Such cases seem to provide interesting and simple
examples of time-dependent backgrounds where nonperturbative instabilities such as
bubbles of nothing \bubbles\
can be the leading effects.\foot{
A similar situation for time-independent orbifolds was found in \aharony.}

\bigskip
\bigskip
\centerline{\bf Acknowledgements}
\medskip

We would like to thank Allan Adams,   Ji\v{r}\'{\i} Bi\v{c}\'{a}k, Steve Giddings,
Matthew Headrick, Simeon Hellerman, Matthew Kleban, Albion Lawrence,
Xiao Liu, Liam McAllister,
Lubo\v s Motl,
Mohammad Sheikh-Jabbari,
 Steve Shenker, and Eva Silverstein for very useful and helpful discussions.
J.M. is grateful to the Harvard Theory Group for hospitality.
The work of M.F. was supported by the DOE under contract
DE-AC03-76SF00515, by the A.P. Sloan Foundation, and
by a Stanford Graduate Fellowship.
The work of J.M. was supported in part by National Science
Foundation grant PHY00-97915.

\listrefs

\bye